\documentclass[reprint,aps,pra,twocolumn,nofootinbib]{revtex4-2}
\pdfoutput=1
\usepackage[utf8]{inputenc}
\usepackage[english]{babel}
\usepackage[T1]{fontenc}
\usepackage{amsmath}
\usepackage{amssymb}
\usepackage[colorlinks]{hyperref}

\usepackage{float}
\makeatletter
\let\newfloat\newfloat@ltx
\makeatother

\usepackage{algorithm}
\usepackage[noend]{algpseudocode}
\usepackage{graphicx}
\usepackage{subfig}
\usepackage[encapsulated]{CJK}

\usepackage{lipsum}

\begin{document}
\begin{CJK*}{UTF8}{gbsn}
\title{Investigating quantum approximate optimization algorithms \\under bang-bang protocols}

\author{Daniel Liang}
\affiliation{X, The Moonshot Factory, Mountain View, California 94043, USA}
\affiliation{Quantum Information Center, University of Texas at Austin, Austin, Texas 78712, USA}
\author{Li Li (李力)}
\affiliation{Google Research, Mountain View, California 94043, USA}
\author{Stefan Leichenauer}
\affiliation{X, The Moonshot Factory, Mountain View, CA 94043, USA}

\begin{abstract}
The quantum approximate optimization algorithm (QAOA) is widely seen as a possible usage of noisy intermediate-scale quantum (NISQ) devices. We analyze the algorithm as a bang-bang protocol with fixed total time and a randomized greedy optimization scheme. We investigate the performance of bang-bang QAOA on MAX-2-SAT, finding the appearance of phase transitions with respect to the total time. As the total time increases, the optimal bang-bang protocol experiences a number of jumps and plateaus in performance, which match up with an increasing number of switches in the standard QAOA formulation. At large times, it becomes more difficult to find a globally optimal bang-bang protocol and performances suffer. We also investigate the effects of changing the initial conditions of the randomized optimization algorithm and see that better local optima can be found by using an adiabatic initialization.
\end{abstract}
\maketitle
\end{CJK*}

\section{Introduction}
The use of quantum computation to solve problems deemed hard for classical computation is an area of massive interest in both the physics and computer science communities. One candidate algorithm for practical speedups on noisy intermediate-scale quantum (NISQ) devices is the quantum approximate optimization algorithm (QAOA) proposed by \citet{farhi2014quantum}. The QAOA involves switching between two Hamiltonians, with the number of switches being defined by a parameter called $p$, as well as an optimization process to control how long each Hamiltonian should be applied.

As stated in the original QAOA paper, the common belief is that $p$ controls the approximation ratio of QAOA, and so $p$ should be as large as possible before the circuit becomes too deep and is overwhelmed by hardware noise~\citep{arute2020quantum_qaoa}. Indeed, \citet{farhi2014quantum} were able to show that as $p \rightarrow \infty$, QAOA is able to achieve a perfect approximation ratio, since in that limit QAOA is as powerful as Adiabatic Quantum Computation~\citep{farhi2000quantum, Kadowaki_1998}.

However, a recent paper from \citet{shaydulin2019evaluating} gave evidence to the contrary, stating that the optimization of variational parameters is difficult at large $p$, and performance improvements at large $p$ are marginal when dealing with bounded computation in the optimization process. We provide further evidence of this. Inspired by \citet{day2019glassy}, we give data from a large-scale classical simulation of a modification to QAOA, which we call bang-bang QAOA, applied to the problem of MAX-2-SAT. While not necessarily practical for NISQ devices, this modification acts as a thought experiment to show that even in the case where $p$ is allowed to be fairly large, while the total time is instead bounded, one does not see large improvements with greater values of $p$. Similar to \citet{shaydulin2019evaluating}, we assert that this is because of a proliferation of local optima, making it difficult to find optima that are close to the global optima.

While we know that as $p\to \infty$ that one can choose the QAOA parameters to correspond to a Trotterized adiabatic quantum computation and achieve a perfect approximation ratio~\citep{farhi2014quantum}, in the finite $p$ regime it is not fully understood whether or not the optimal parameters for QAOA should appear adiabatic \citep{brady2020optimal, PhysRevX.7.021027, bapat2018bangbang, mbeng2019quantum}. We emphasize that while some of these works studied QAOA through the lens of bang-bang control theory, they ultimately focused on applying bang-bang control theory to the adiabatic algorithm and seeing if the result resembles QAOA. In contrast, we assume a bang-bang structure and use a randomized greedy optimization algorithm to study the solution space of bang-bang QAOA algorithms with a fixed total time. In the bang-bang QAOA model, we see that when total time is small, the best protocols do not appear adiabatic, but rather correspond to finite-$p$ implementations of standard QAOA. When the total time is large, the proliferation of local optima means that our optimization procedure depends strongly on the initialization. Though we cannot say much about any global optima, we do see that adiabatic initialization provides a good heuristic for finding better protocols.

\section{QAOA and Bang-bang Protocols}
As its name suggests, the QAOA is a quantum-based algorithm for combinatorial optimization designed to find the set of inputs that approximately optimizes an efficiently computable objective function. At a high level, it does so by encoding this objective function along the diagonal of a Hamiltonian. The algorithm then tries to find a circuit that efficiently brings the state $|0^n\rangle$ as close as possible to the ideal state by applying two different Hamiltonians. We will first describe the standard QAOA as given by \citet{farhi2014quantum}, followed by our bang-bang QAOA modification. Note that there exist a wide variety of other interesting modifications to the standard QAOA \citep{li2019quantum, Hadfield_2019, cook2019relationships}.

\subsection{Standard QAOA}
Let $E = \sum_{x \in \{0, 1\}^n}f(x)|x\rangle \langle x|$ be the Hamiltonian that encodes the objective function $f$ along its diagonal. $E$ will be referred to as the \textit{constraint Hamiltonian} while $X^{\otimes n} = (|0\rangle\!\langle 1| + |1\rangle\!\langle0|)^{\otimes n}$ will be referred to as the \textit{mixing Hamiltonian}. Now let $\beta_1, \cdots \beta_p, \gamma_1, \cdots \gamma_p$ be positive real parameters for QAOA with depth $p$ \footnote{The QAOA parameters are not always restricted to be positive, but the $\beta$ parameters are naturally periodic, and so too are the $\gamma$ parameters when (as will be the case here) the objective function takes on integer values.}.
The state produced by QAOA is then
$$|\psi\rangle = e^{i\beta_p X^{\otimes n}} e^{i \gamma_p E} \cdots e^{i\beta_1 X^{\otimes n}} e^{i \gamma_1 E}H^{\otimes n}|0^n \rangle,$$
where $H = (|0\rangle\!\langle 0| + |1\rangle\!\langle0|+|0\rangle\!\langle 1|-|1\rangle\!\langle 1|)/\sqrt{2}$ is the Hadamard operator. The $2p$ parameters are optimized based on the expectation value $\langle \psi |E|\psi\rangle$ in order to increase the chance of measuring a good input when $|\psi\rangle$ is measured in the computational basis.

\subsection{Bang-bang QAOA}\label{sec:bangbang}

\begin{figure}[hbt!]
\includegraphics[width=\columnwidth]{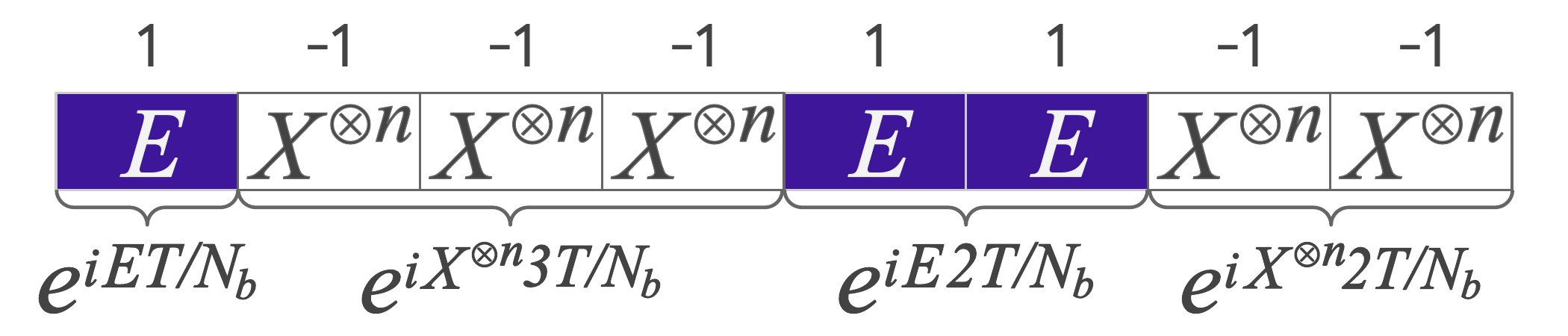}
\caption{\raggedright Example of a bang-bang QAOA protocol with $N_b = 8$. It is equivalent to a $p = 2$ QAOA protocol with $\gamma_1 = T/N_b$, $\beta_1 = 3T/N_b$, $\gamma_2 = 2T/N_b$, and $\beta_2 = 2T/N_b$. The protocols are read from left-to-right in order of applying the Hamiltonians.}
\label{fig:bangbang_example}
\end{figure}

A bang-bang control scheme is a system that switches abruptly between two different modes and is an important part of optimal control theory \citep{pontryagin1986optimal}. Here, the two modes will be the application of the Hamiltonians $E$ and $X^{\otimes n}$ respectively. In order to explore the space of protocols computationally, we break up the total time $T$ into $N_b$ blocks of time of $T/N_b$ time each, with each block assigned to one or the other Hamiltonian. A bang-bang QAOA protocol then involves iterating through the blocks applying the corresponding Hamiltonian for $T/N_b$ amount of time. This simply involves applying either $e^{iET/N_b}$ or $e^{iX^{\otimes n}T/N_b}$, respectively. See Fig.~\ref{fig:bangbang_example} for an example.

If one were to translate a bang-bang QAOA protocol into the language of the standard QAOA, the $p$ value of said protocol could be as large as $N_b/2$. However, the total amount of time is at most $T$. In addition, in the large $N_b$ limit this bang-bang QAOA model can approximate any standard QAOA protocol such that $\sum_{i=1}^p (\beta_i + \gamma_i) \approx T$. Later on we will also argue that it is not worthwhile to consider QAOA protocols for large $T$ or $\sum_{i=1}^p (\beta_i + \gamma_i)$ in the bang-bang and standard QAOA, respectively, due to the difficulty of optimization.

\section{MAX-2-SAT}

For boolean expressions a conjunction is a logical AND and is typically represented as $\wedge$. A disjunction is a logical OR that is represented as $\vee$. Finally, $\neg$ as a unary operator represents logical negation. Given boolean values $x_0, \cdots x_{n-1}$, a $k$-CNF (Conjuctive Normal Form) is the conjunction over disjunctive clauses of size $k$. More intuitively, a $k$-CNF is an AND-of-ORs where each OR involves $k$ boolean values. 2-SAT is then the problem of determining if there exists a assignment of $x_0, \cdots x_{n-1}$ such that a given 2-CNF is satisfied. The natural optimization version of the problem, MAX-2-SAT, is then the problem of determining the maximum number of clauses satisfiable by an assignment of $x_0, \cdots x_{n-1}$.

It is important to make the distinction between 2-SAT and MAX-2-SAT; 2-SAT is in \textsc{P} \citep{doi:10.1002/malq.19670130104, ASPVALL1979121, doi:10.1137/0205048} while MAX-2-SAT is \textsc{NP}-Hard \citep{GAREY1976237}. Hardness of approximation results have shown that no Polynomial-Time Approximation Schemes (PTAS) exist for MAX-2-SAT with approximation ratios better than $\frac{21}{22} \approx 0.955$ \citep{Hastad:2001:OIR:502090.502098} assuming $\textsc{P} \neq \textsc{NP}$ and $\sim0.943$ \citep{Khot:2007:OIR:1328722.1328735} when also assuming the unique games conjecture. Here the approximation ratio of an algorithm refers to a guarantee that an algorithm with approximation ratio $r$ for a problem instance with optimal solution $C_{max}$ achieves a result of at least $r C_{max}$ (potentially only with high probability if randomized or quantum). There does, however, exist an efficient algorithm based on semi-definite programming that achieves an approximation ratio of $0.94$ \citep{Lewin2002ImprovedRT}. It is worth noting that a uniformly random assignment of literals will satisfy $\frac{3}{4}$ of the clauses in expectation. By the probabilistic method this also ensures that at least $\frac{3}{4}$ of the clauses are always satisfiable.

Finally, we will show how to encode MAX-2-SAT as a Hamiltonian $E$. Given a disjunctive clause of 2 literals, there is exactly one assignment that does not satisfy the clause. We then design a diagonal Hamiltonian for the clause that is $1$ for every literal assignment where the clause is not satisfied. For instance, given the clause $\mathcal{C} = (x_i \vee \neg x_j)$ where $i < j$ the only assignment that does not satisfy $\mathcal{C}$ is $x_i = 0, x_j = 1$, where $1$ is True and $0$ is False. We can then define the Hamiltonian for $\mathcal{C} = (x_i \vee \neg x_j)$
\begin{equation}
    E_\mathcal{C} = I^{\otimes n} - I^{\otimes i-1}\otimes |0 \rangle \langle 0| \otimes I^{\otimes j-i-1} \otimes |1 \rangle \langle 1| \otimes I^{n-j}.
\end{equation}
The diagonal of this Hamiltonian is then $1$ for every computational basis state $x$ such that $x_i \neq 0$ or $x_j \neq 1$. The Hamiltonian of MAX-2-SAT is then the sum over all Hamiltonians induced by the clauses $\{\mathcal{C}\}$ in the 2-CNF
\begin{equation}\label{eq:expected_clauses}
    E = \sum_{\mathcal{C}} E_\mathcal{C}.
\end{equation}
One can see that the diagonal encodes the number of clauses satisfied by the assignment of literals.
The objective function of QAOA is the expected number of satisfied clauses, $\langle\psi|E|\psi\rangle$.
If $C_{max}$ is the maximum number of satisfiable clauses, by linearity of expectation this leads to the expected approximation ratio
\begin{equation}\label{eq:obj}
    f_{obj} = \frac{\langle\psi|E|\psi\rangle}{C_{max}}.
\end{equation}
While we cannot normally directly compute the approximation ratio without knowing $C_{max}$, maximizing $\langle\psi|E|\psi\rangle$ will also maximize $f_{obj}$ due to being related by a constant factor.

\section{Methods}
In this section we will simply outline preliminary information to understand our results. We will henceforth set the number of variables in our MAX-2-SAT instances to be $10$. This is due to the exponential nature of the dimension of the Hilbert space with respect to the variables (i.e. qubits). This means that adding a single variable will double the amount of computation needed.

\subsection{Stochastic Descent}
QAOA optimizes the quantum circuit in order to increase the probabilities of getting a good measurement. Given a bang-bang QAOA protocol $P$ that produces state $|\psi_P\rangle$, our objective function $f_{obj}(P)$ will be the expected approximation ratio of the resulting state in Eq.~\ref{eq:obj}. In the bang-bang QAOA our protocols fall into a discrete space. As such, we use the following greedy randomized optimization approach introduced by \citet{day2019glassy} with $k=1$:

\begin{algorithm}[hb]
   \caption{Stochastic descent ($\mathrm{SD}_k$)}
\begin{algorithmic}[1]
\State \textbf{Input:} $N_b$, $T$, $k$
\State \textbf{Routines:} RandomProtocol, FindAllUpdate, RandomShuffle, UpdateProtocol, $f_{obj}$
\State \emph{initialize:}
\Statex{~~~//Initialize protocol at random}
\State ~~~ $P_{old}\gets$ RandomProtocol($N_b$)
\Statex{//Finds the list of updates with at most $k$-flips}
\State ListOfAllUpdates~$\gets$~FindAllUpdate($N_b$, $k$)
\State \emph{shuffle:}
\Statex{~~~//Shuffle updates in a random order}
\State ~~~ListOfAllUpdates~$\gets$~RandomShuffle(ListOfAllUpdates)
\Statex{//Iterate over all possible update}
\For{$update$ \textbf{in} ListOfAllUpdates}
\Statex{~~~//Update protocol given the specified update}
\State $P_{new} \gets \text{UpdateProtocol}(P_{old}, update)$
\Statex{~~~//Evaluates the objective function of each protocol\\~~~and compares them}
\If{$f(P_{new}) > f(P_{old})$}
\State $P_{old} \gets P_{new}$
\Statex{~~~~~~~~~//If update accepted, then restart for loop}
\State \textbf{goto} \emph{shuffle}
\EndIf
\EndFor
\Return $P_{old}$
\end{algorithmic}
\end{algorithm}

If bang-bang protocols are viewed as bit strings, the algorithm randomly iterates through all protocols of Hamming distance at most $k$ away from the current one and updates itself to the first protocol it finds that performs better. If no protocol is better, then we say that the current protocol is a $k$-local optimum. Note that the number of protocols that need to be considered at each update grows with $k$ as $\sum_{i=1}^k \binom{N_b}{i}$, which is near exponentially for $k \leq N_b / 2$. Thus, increasing $k$ quickly becomes very computationally expensive.

\subsection{Random Protocol Initialization \label{ssec:initialization}}
An interesting aspect of Stochastic Descent ($\mathrm{SD}$) is the distribution with which the initial random protocol is drawn from. The original algorithm proposed by \citet{day2019glassy} uniformly samples at random. In this paper, we propose two new initialization to study the relation between bang-bang QAOA with Adiabatic Quantum Computation.

We define $\xi(q)$ to be the Bernoulli random variable with probability $q$ of being $1$ and $1-q$ of being $-1$. Varying $q$ as a function over blocks generates three different random initialization methods:

\begin{description}
	\item[Adiabatic \label{item:adiabatic}] $\prod_{i=1}^{N_b} \xi(i/N_b)$. It favors $X^{\otimes n}$ in early blocks and $E$ in late blocks of the protocols.
	\item[Uniform \label{item:uniform}] $\prod_{i=1}^{N_b} \xi(0.5)$. The probabilities of $X^{\otimes n}$ and $E$ being sampled are equal. This is the default distribution used in this paper.
	\item[Antiadiabatic \label{item:anti_adiabatic}] $\prod_{i=1}^{N_b} \xi(1-i/N_b)$. It favors $E$ in early blocks and $X^{\otimes n}$ in late blocks of the protocols.
\end{description}

\subsection{Correlator} \label{par:correlator}
Given a set $S$ of protocols with $N_b$ blocks, we define the correlator  of $S$ as the following: View a protocol as a collection of values $P \in \{-1, 1\}^{N_b}$, where $P_i$ refers to the value at block $i$. Let $\overline{P_i} = \frac{1}{|S|}\sum_{P \in S}P_i$ represent the empirical average of block $i$ over all protocols in the set $S$. The correlator  is defined as a certain variance of the protocol values,
\begin{equation}
    \sigma=\frac{1}{N_b |S|}\sum_{i=1}^{N_b} \sum_{P \in S} (P_i - \overline{P_i})^2=1 - \frac{1}{N_b}\sum_{i=1}^{N_b} \overline{P_i}^2.
\end{equation}

A small correlator  means the protocols in $S$ are similar to each other. Intuitively, it is really an ``anti-correlator'' as the value is small when the protocols are similar. We choose to keep the same name as \citet{day2019glassy} for consistency.

\subsection{Protocol Smoothing \label{ssec:protocol_smoothing}}
It is also important to analyze the actual structure of bang-bang QAOA protocols after Stochastic Descent. While one could plot the protocols themselves as $\{-1, 1\}$ values along a time-scale, this does little to see the effects of how a protocol may favor one Hamiltonian over the other at different points in time. As such, we also opt to smooth the protocols by taking a rolling average. In addition, this smoothing allows us to properly see how close many of these bang-bang QAOA protocols are to being standard QAOA protocols for small total time by smoothing over minor deviations.

More formally, let $w$ be a positive integer known as the \textit{window size}. Similar to the correlator , we will view bang-bang QAOA protocols as $P \in \{-1, 1\}^{N_b}$ where $P_i$ refers to the value at block $i$. We then defined the \textit{smoothed protocol} $$P' \in [-1, 1]^{N_b - w + 1}$$ where $$P'_i = \frac{1}{w}\sum_{j = 0}^{w-1} P_{i + j}.$$

\subsection{Problem Instances}
The $2$-DNFs used were constructed by randomly generating a clause with two unique indices drawn uniformly, as well as whether or not to negate each variable. Several of these clauses are then independently created, with the number of random clauses $n_c$ being a parameter specified at runtime. Note that it is possible that two identical clauses are created, and by the Birthday Paradox we expect this to happen when $n_c \gtrsim n$. While this is the regime that we end up creating our problem instances with, one can simply repeat the process an expected constant number of times until success. We ensure that there are no identical clauses in our problem instances. See Appendix~\ref{appendix:problem_instance} for the actual problem instances used of $10$, $20$, and $30$ clauses respectively. For clarity, we focus on the $10$ clause problem instance in the proceeding results section.

\section{Results}
\begin{figure*}[hbt!]
  \includegraphics[width=\textwidth]{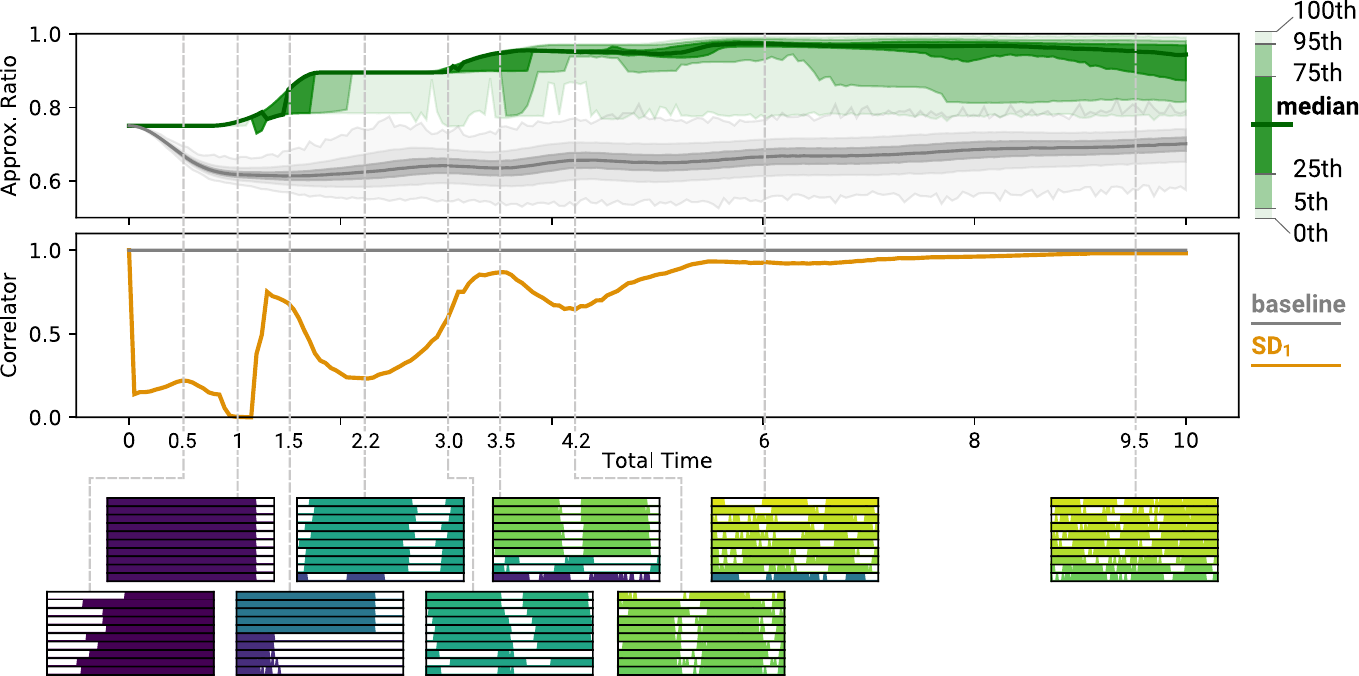}
  \caption{\raggedright Top panel represents aggregate statistics on the expected approximation ratio with respect to total time $T$. Percentiles are given respectively. 10,000 protocols are sampled per time-step with $N_b = 200$ blocks. Fill grey region represents protocols without $\mathrm{SD}_1$ applied. Shaded regions for each color denote the percentile regions as indicated in the legend. The initial protocols are sampled uniformly at random. The bottom panel contains the corresponding correlator. The vertical lines and thumbnails indicate 10 randomly sampled protocols found by $\mathrm{SD}_1$ at corresponding total times. More details of the profiles of protocols are in Figs.~\ref{fig:p_approx} and~\ref{fig:protocol_smooth_compare}.
  In the text we discuss the features of this figure, such as the rapid increase in median expected approximation ratio at specific times and the increase in the correlator  over time.}
  \label{fig:10_200_random}
\end{figure*}

In Fig.~\ref{fig:10_200_random} we can see how the protocols drawn uniformly at random perform without being optimized with $\mathrm{SD}$ (in grey) as compared to $\mathrm{SD}_1$ (in green), with a substantial increase in expected approximation ratio. Even with $\mathrm{SD}_1$ it is easy to see the benefit of a greedy optimization strategy for bang-bang QAOA. Interestingly, without $\mathrm{SD}_1$, protocols perform worse than the naive classical algorithm of a uniformly random assignment of variables, which achieves at least a $\frac{3}{4}$ approximation ratio.

\subsection{Small time regime}

We will refer to the small time regime as $T \lesssim 6$, though this value is likely problem instance specific. The important aspect of these figures in this regime is the rapid increase in median expected approximation ratio around $T \approx 1.5$ and $T \approx 3.5$, which we refer to as a \textit{phase transition} in performance. We attribute this to be the minimal total time needed for protocols to start enacting non-trivial behavior, corresponding to $\mathrm{SD}_1$ converging on a $p \approx 2$  and $p \approx 3$ protocol respectively when viewed as a standard QAOA protocol.

\begin{figure}[hbt!]
\includegraphics[width=\columnwidth]{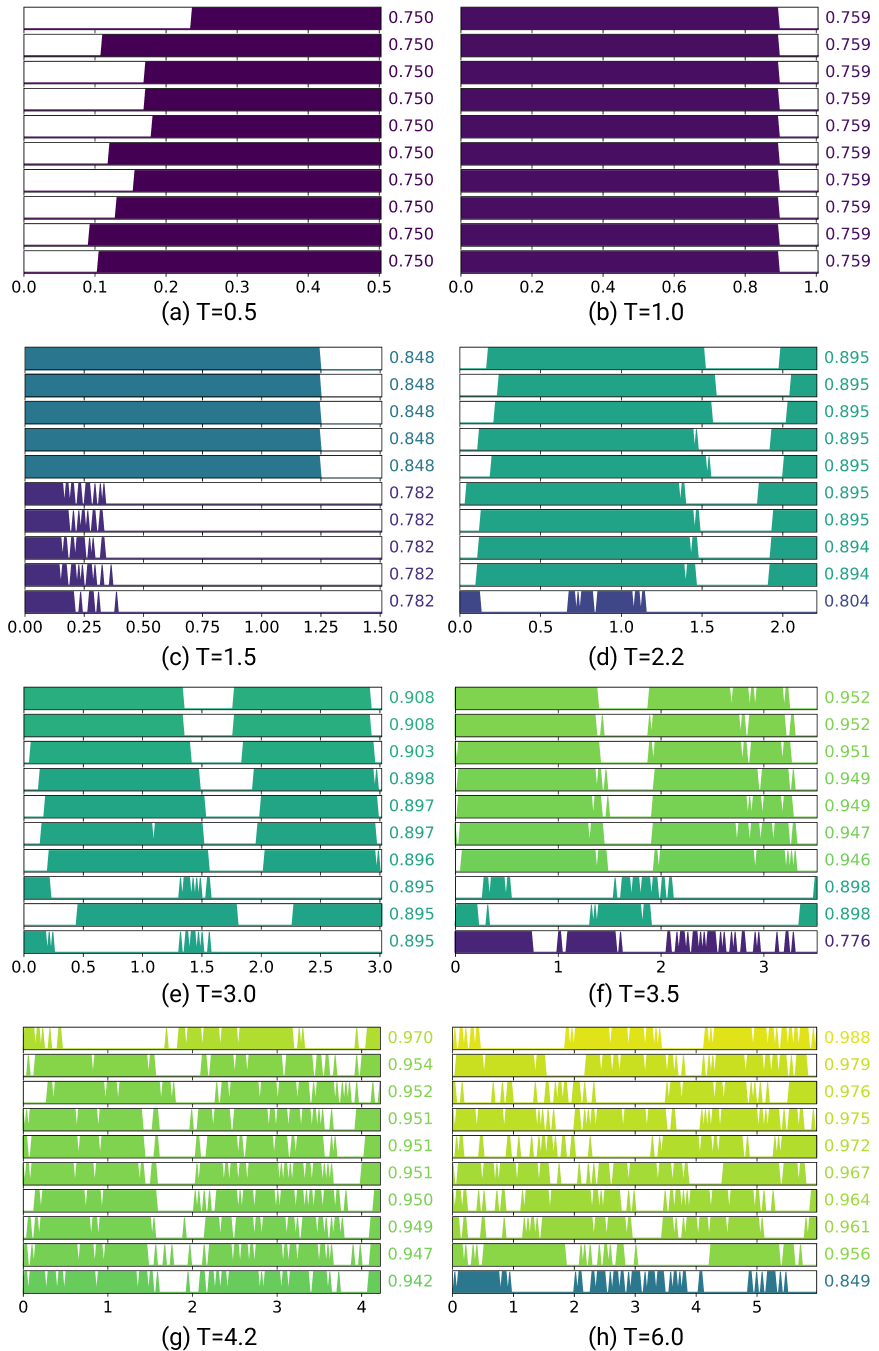}\label{fig:protocols_small_time}
\caption{\raggedright Illustrates how bang-bang QAOA protocols converge towards standard QAOA protocols with small $p$ values for $N_b = 200$ and $10$ clauses with uniform random initialization. Drawing $10$ protocols uniformly at random with a window size $w = 1$. (a) $T = 0.5$ and (b) $T = 1.0$ are before the first transition point and are similar to standard QAOA with $p=1$. (c) $T = 1.5$ is during the transition, where the rapid increase in the correlator  is due to the mixture of two kinds of protocols. (d) $T = 2.2$ is after the transition where protocols resemble $p=2$. Likewise, (e) $T = 3.0$ is before the second transition, (f) $T = 3.5$ is during, and (g) $T = 4.2$ is after. (h) $T = 6.0$ is at the point where the median expected approximation ratio begins to plateau. Colormap is based on the expected approximation ratios of protocols and the numbers on the right indicate their corresponding values. See Fig.~\ref{fig:10_200_random} to see the transition points.}
  \label{fig:p_approx}
\end{figure}

We can see from Figs. \hyperref[fig:protocols_small_time]{3(a)} and \hyperref[fig:protocols_small_time]{3(b)} that at very small time the protocols only apply each Hamiltonian once. Then as the total time increases, the protocols then transition into two switchbacks as seen in Figs.~\hyperref[fig:protocols_small_time]{3(d)} and \hyperref[fig:protocols_small_time]{3(e)} leading to the median expected approximation ratio to increase substantially. This again repeats with $p \approx 3$ like in Fig.~\hyperref[fig:protocols_small_time]{3(f)}, however the increase in median expected approximation ratio is not as great as before.

Looking at the correlator  in Fig.~\ref{fig:10_200_random}, at $T = 0$ it starts off around $1$ since every starting protocol is a local optimum with a uniform probability of being selected. As $\mathrm{SD}_1$ begins to optimize towards specific protocols, the value then quickly drops as there are only a few and very similar local optima. It is then when transitioning to a new local optima that the correlator  begins to temporarily spike, as there is a mixture of protocols as with Fig.~\hyperref[fig:protocols_small_time]{3f}. Once the transition has finished, the correlator  then quickly decreases again. However, there is a general trend towards protocols becoming uncorrelated as the number of local optima start increasing with total time.

\subsection{Large time regime}\label{sec:large_time}

It is then at large time that increasing total time no longer becomes as beneficial for the global optima and the number of local optima starts to increase rapidly. Here, despite the fact that the best protocols continue to do marginally better at large $T$, looking carefully at Fig.~\ref{fig:10_200_random} the median trends downward. We believe this is due to the local optima no longer being close to the global as it becomes more and more difficult to find better optima using $\mathrm{SD}_1$. This tells us that within the realms of greedily optimized bang-bang QAOA, there is more than enough time necessary for a near-optimal protocol and any extra total time contributes to extra degrees of freedom that make optimization more difficult. This becomes even more apparent as the number of clauses increases and the median protocol begins to quickly fall off as total time increases.

\begin{figure}[hbt!]
\includegraphics[width=\columnwidth]{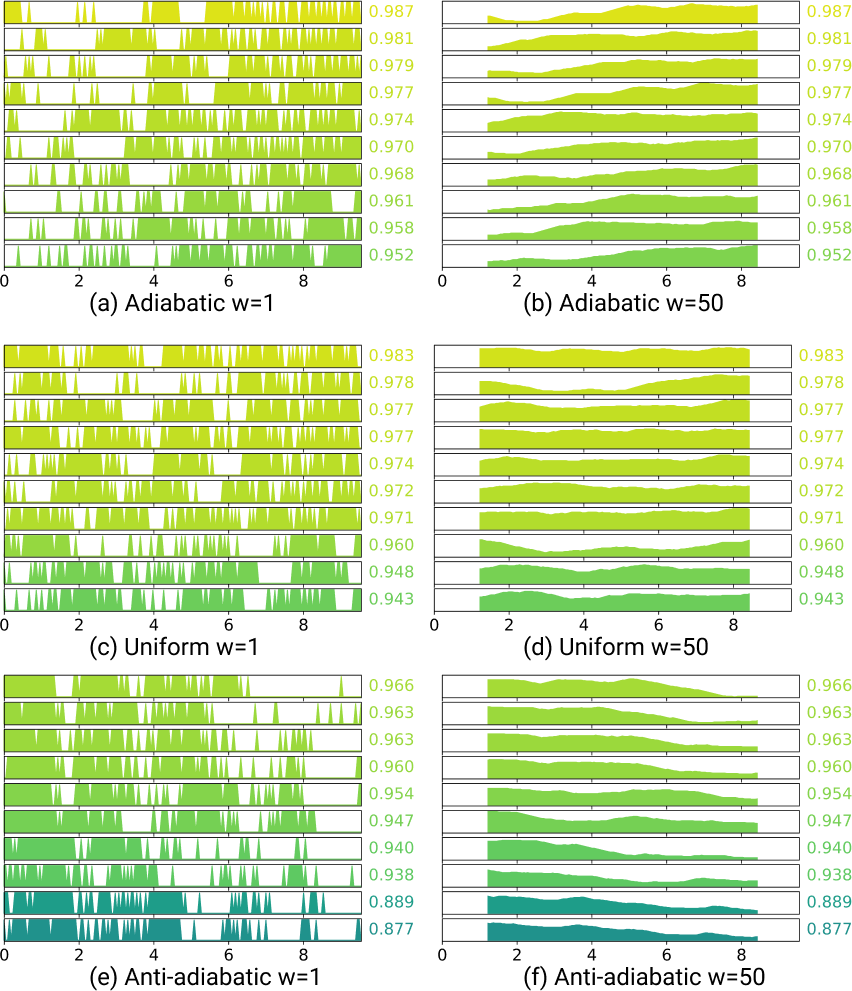}

  \caption{\raggedright Demonstrates the profiles of local optima at large time ($T=9.5$) with [(a) and (b)] \hyperref[item:adiabatic]{adiabatic}, [(c) and (d)] \hyperref[item:uniform]{uniform}, and [(e) and (f)] \hyperref[item:anti_adiabatic]{antiadiabatic} initialization for $N_b = 200$ and $10$ clauses.
  Protocols are sampled randomly and smoothed with a window size $w = 1$ and $w=50$ respectively. 
  Within each subfigure, there is no qualitative difference in the profiles between the best and worst protocols and no strong underlying standard QAOA-like structure behind the protocols like with Fig.~\ref{fig:p_approx}. Additionally, the shape of the \hyperref[ssec:protocol_smoothing]{smoothed protocols} being similar to the expected initial protocol even after $\mathrm{SD}_1$ indicates that local optima can be found with qualitatively different protocols. Colormap is based on the expected approximation ratios of protocols and the numbers on the right indicate their corresponding values. Note that the same $10$ protocols are shown with different smoothing for each initialization. Initialization explained in Sec.~\ref{ssec:initialization}.}
  \label{fig:protocol_smooth_compare}
\end{figure}

Looking at the protocols in Fig.~\hyperref[fig:protocol_smooth_compare]{4(c)} without smoothing, there is no apparent discernible profile with the protocols in how they relate to their expected approximation ratio. However, we do see in Figure \hyperref[fig:protocol_smooth_compare]{4d} that the protocols tend to favor the constraint Hamiltonian and appear neither adiabatic nor antiadiabatic. Together with Figs. \hyperref[fig:protocol_smooth_compare]{4(b)} and \hyperref[fig:protocol_smooth_compare]{4(f)}, we find the trend of protocols remain qualitatively similar to their initialization.

\subsection{Summary}
Below a certain total time $T$, no bang-bang QAOA protocol does well since the resulting unitary of the circuit will still be close to identity. After a certain point, a select few protocols start exhibiting nontrivial behavior, which is then found by $\mathrm{SD}_1$ and the protocols transition to a much better expectation. The protocols do not initially benefit from further increases in $T$, which causes performance to plateau. Then at some point, $T$ becomes large enough to allow for another set of non-trivial behavior, and this process continues until the transition to large time. At this point $T$ becomes too large and it becomes too difficult to find a solution near the global optima. The protocols then start exhibiting less structure and become very different from each other quantitatively based on the correlator, but qualitatively do not deviate far from their initialization.

\subsection{100 vs 200 blocks}
\begin{figure}[hbt!]
\includegraphics[ width=0.95\columnwidth]{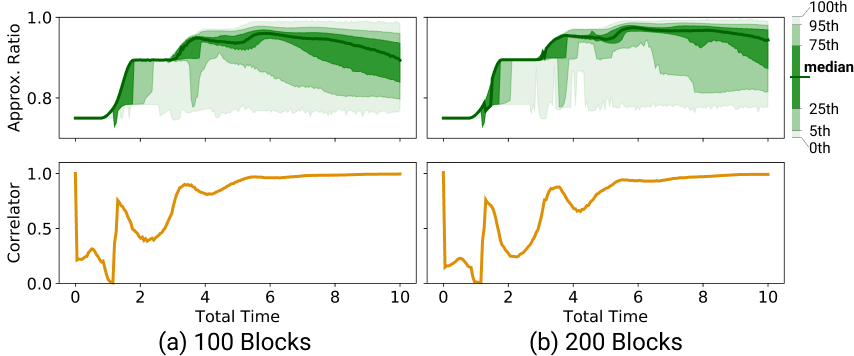}
  \caption{\raggedright (a) 100 vs (b) 200 blocks are compared side-by-side for the same problem instance containing 10 clauses. With 200 blocks one sees slightly better median expected approximation ratios.}
  \label{fig:blocks}
\end{figure}

In Fig.~\ref{fig:blocks} we illustrate how the number of blocks $N_b$ affects the expected approximation ratio of bang-bang QAOA. While the shape of both graphs are very similar, one can see that between Figs. \hyperref[fig:blocks]{5(a)} and \hyperref[fig:blocks]{5(b)} $N_b = 200$ tends to give better expected approximation ratios, especially when the total time $T$ becomes large. Additionally, the correlator  dips lower around the phase transitions, indicating that the protocols actually concentrate better with larger $N_b$ around the phase transitions, despite the fact that there are exponentially more protocols available as $N_b$ increases.

\subsection{Larger number of clauses}

\begin{figure}[hbt!]
\includegraphics[ width=0.95\columnwidth]{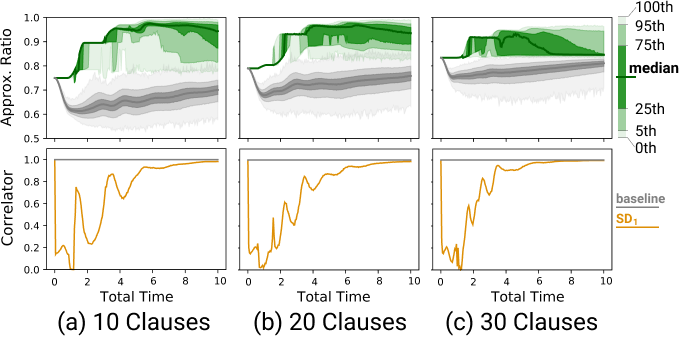}
  \caption{\raggedright Expected approximation ratio and the correlator when run on the (a) \hyperref[ssec:10_clause]{$10$}, (b) \hyperref[ssec:20_clause]{$20$} and (c) \hyperref[ssec:30_clause]{$30$ clauses} problem instances. 10,000 protocols are sampled per time-step with $N_b = 200$ blocks. Grey plots represent protocols without $\mathrm{SD}_1$ applied. The initial protocols are sampled uniformly at random. Because the ratio of satisfiable clauses tends to decrease as the total number of clauses increases, the baseline of random guessing is able to perform better as it is always able to satisfy $\frac{3}{4}$ of all clauses.}
  \label{fig:aggregate_plots_clauses}
\end{figure}

It is of course important to analyze more than a single problem instance. In Fig.~\ref{fig:aggregate_plots_clauses}, we find that the overall behavior remains relatively consistent between the problem instances with $10$, $20$, and $30$ clauses respectively. More specifically, we see that the median expected approximation ratio increases in jumps in the small time regime, before trailing off at large time. This decay in median expected approximation ratio is especially pronounced in Figure \hyperref[fig:aggregate_plots_clauses]{6c}. Similarly, the correlator moves up and down in the small time regime, though increasing to a value of nearly $1$ as time increases.

\subsection{Effects of Random Initialization} \label{sec:initialization}

\begin{figure}[hbt!]
\includegraphics[ width=\columnwidth]{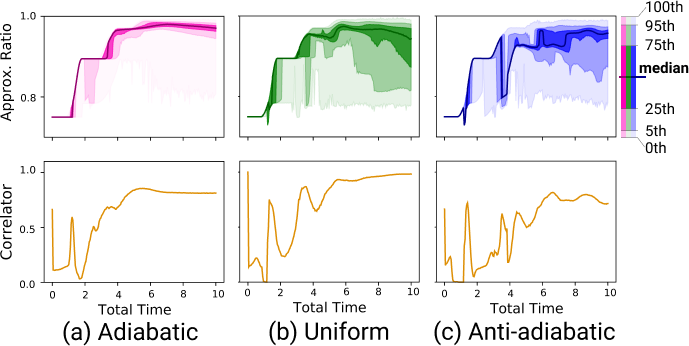}
  \caption{\raggedright Expected approximation ratio and the correlator when using (a) \hyperref[item:adiabatic]{adiabatic}, (b) \hyperref[item:uniform]{uniform}, and (c) \hyperref[item:anti_adiabatic]{antiadiabatic} initialization. The probability distribution that the initial protocol is drawn from does have minor small $T$ and more pronounced large $T$ affects on how the protocols perform, even after $\mathrm{SD}_1$. Here $10$ clauses are used with $N_b = 200$.}
  \label{fig:initialization}
\end{figure}

\begin{figure}[hbt!]
\includegraphics[ width=\columnwidth]{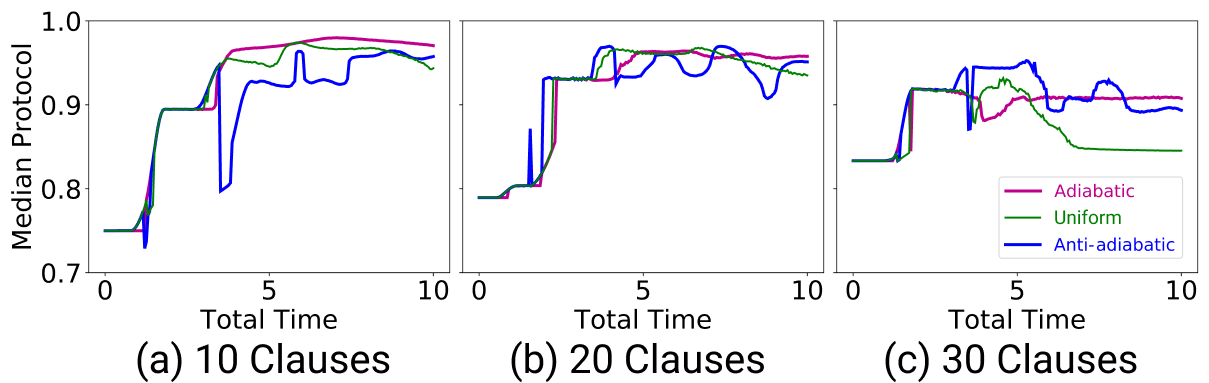}
  \caption{\raggedright Comparison of median expected approximation ratio of protocols based on \hyperref[ssec:initialization]{initialization}. (a) \hyperref[ssec:10_clause]{$10$ clause}, (b) \hyperref[ssec:20_clause]{$20$ clause}, and (c) \hyperref[ssec:30_clause]{$30$ clause} problem instances are presented. Adiabatic initialization tends to consistently do well even at large total time, as opposed to uniform random which tends to drop off. Antiadiabatic initialization leads to high variance in the median expected approximation ratio.}
  \label{fig:initialization2}
\end{figure}

Looking at Fig.~\ref{fig:initialization}, they all perform similarly at small $T$. This is to be expected as there are few local optima such that the initialization only effects the starting distance to the global. However, at large $T$, Fig.~\ref{fig:initialization2} shows us that the uniform random initialization tends to do poorly with respect to the median protocol. The \hyperref[item:adiabatic]{adiabatic initialization} however tends to consistently do well at large $T$, while the \hyperref[item:anti_adiabatic]{antiadiabatic initialization} exhibits large variance in its median expected approximation ratio over time (see Sec.~\ref{ssec:initialization} for the definition of adiabatic and antiadiabatic initialization). As $T$ becomes large, the adiabatic theorem, the driving force behind adiabatic quantum computation \citep{farhi2000quantum}, starts becoming relevant. If intuition from the adiabatic theorem and adiabatic quantum computation extends to the local optima found by $\mathrm{SD}_1$, the local optima found around the adiabatic initialization are then likely to perform better on average than those initialized from uniform or antiadiabatic distributions.

\begin{figure}[hbt!]
  \includegraphics[ width=0.95\columnwidth]{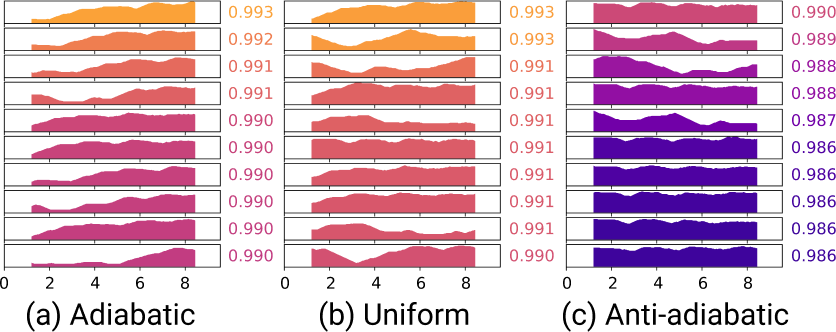}
  \caption{\raggedright Top ten bang-bang QAOA protocols using (a) \hyperref[item:adiabatic]{adiabatic}, (b) \hyperref[item:uniform]{uniform}, and (c) \hyperref[item:anti_adiabatic]{antiadiabatic} initialization at large time ($T=9.5$), \hyperref[ssec:protocol_smoothing]{smoothed} with window size $w = 50$ for $N_b = 200$ with $10$ clauses. Colormap is based on the expected approximation ratios of protocols and the numbers on the right indicate their corresponding values. Since $1$ corresponds to the objective Hamiltonian, an adiabatic protocol will gradually increase in value.}
  \label{fig:50_smooth}
\end{figure}

Further evidence of this can be seen by once again examining the protocols themselves. Looking at Fig.~\ref{fig:protocol_smooth_compare}, we see that the randomly drawn protocols from each initialization are very similar to their expected starting protocol. Finally, if we instead look at the best protocols of each initialization, Fig.~\ref{fig:50_smooth} shows the best protocols appear to be qualitatively more adiabatic: adiabatic initialization leads to strongly adiabatic algorithms, antiadiabatic is largely uniform, and random initialization slightly favors adiabaticity. Thus even though the local optima themselves seem to be unbiased, the best protocols seem to be found in the space around qualitatively more adiabatic protocols than the initialization.

\subsection{Iterations plots}
\begin{figure}[hbt!]
\includegraphics[ width=\columnwidth]{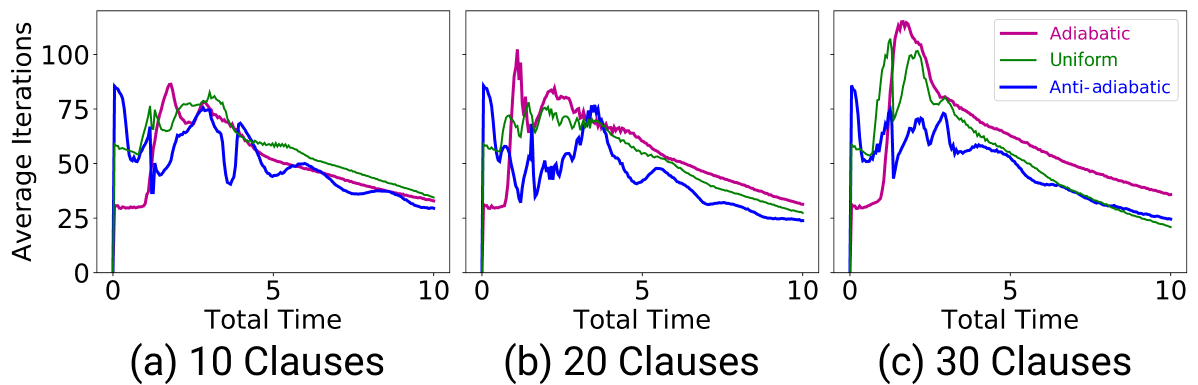}
  \caption{\raggedright Shows the average number of iterations of $\mathrm{SD}_1$ are needed before a local optima are found on the (a) \hyperref[ssec:10_clause]{$10$ clause}, (b) \hyperref[ssec:20_clause]{$20$ clause}, and (c) \hyperref[ssec:30_clause]{$30$ clause} problem instances. As the number of local optima increase with total time, it becomes easier to find one such that the number of iterations quickly decreases.}
  \label{fig:iterations}
\end{figure}

As a final demonstration of the difficulty of finding the globally optimal protocol as the number of local optima increases with $T$, we examine the average number of iterations $\mathrm{SD}_1$ needs to find a local optima. Looking at Fig.~$\ref{fig:iterations}$, we can see that the number of iterations needed increases at the first phase transitions, before decaying as $T$ increases. This is true for all three initialization. What this effectively means is that the distance from a random protocol to its nearby local optima decreases with $T$ regardless of the three starting positions.

\section{Conclusion}
Ultimately, it is not clear the bang-bang QAOA should be used in practice with NISQ devices. As stated in Sec.~\ref{sec:bangbang}, the depth of the circuit can potentially be as large as $N_b / 2$, which is exactly the reason $p = O(1)$ is used in standard QAOA to avoid this. However, as a thought experiment as to the value of the $p$ parameter itself, this serves as further evidence that larger $p$ values are not necessary to achieve the best approximation ratios when the optimization process is limited to bounded computation. We see that while a minimal amount of time is needed for bang-bang QAOA protocols to achieve non-trivial approximation ratios, they fail to substantially improve in the median expected approximation ratio for larger $T$. It is also not clear from the data alone how good of an approximation ratio one can get using bang-bang QAOA efficiently.

Due to the nature of classical simulation of quantum mechanics, collecting data is incredibly time intensive even with parallelization of sample collection. For example, because $\mathrm{SD}_k$ takes time exponential in $k$ for small $k$, we were restricted to $\mathrm{SD}_1$. Additionally, the number of variables was only set to $10$, creating very small 2-SAT instances. It will be interesting to see if these behaviors remain the same even with larger problem instances and/or using $\mathrm{SD}_k$ for $k > 1$. Additionally, though we examine $N_b = 100$ and $N_b = 200$ in Fig.~\ref{fig:blocks} there is not currently enough data to draw strong conclusions between the relationship between $N_b$ and performance.

Another consideration is that various other modifications to QAOA such as \citet{li2019quantum}, which modifies the objective function, can be combined with bang-bang QAOA. One compelling modification could involve the ability to apply a Hamiltonian for negative time, corresponding to negative $\{\beta_i\}$ and $\{\gamma_i\}$ parameters in standard QAOA such that total time becomes $T = \sum_i |\beta_i| + |\gamma_i|$ \citep{farhi2014quantum_maxcut}. Changes to $\mathrm{SD}$ are necessary, such as redefining the distance metric between protocols beyond Hamming Distance, as well as preventing the cancellation of Hamiltonians. How these modifications work in tandem with bang-bang QAOA may lead to interesting phenomena that could potentially lead to a more practical algorithm.

Open source code of this paper is available in github repository~\cite{code}.

\begin{acknowledgments}
The authors thank Zan Armstrong for suggestions in data visualization, and Murphy Yuezhen Niu, John Platt for their review and comments.
Daniel Liang would like to thank the Simons It from Qubit Collaboration and Dr. Scott Aaronson for supporting him.
X, formerly known as Google[x], is part of the Alphabet family of companies, which includes Google, Verily, Waymo, and others~\cite{googlex}. Quantum simulation and $\mathrm{SD}$ in this paper were implemented using Cirq~\cite{cirq} and Apache Beam~\cite{apache_beam}.
\end{acknowledgments}

\appendix
\onecolumngrid

\section{Problem Instances Used in This Paper}
Notation: $\vee$ represents a disjunction, $\wedge$ represents a conjunction, and $\neg$ represents logical negation.
\label{appendix:problem_instance}

\subsection{10 Clauses \label{ssec:10_clause}}
\begin{align*}
f(x) =
& (\neg x_8 \vee x_9) &&  \wedge && (\neg x_5 \vee x_7) && \wedge && (x_0 \vee \neg x_6) && \wedge && (\neg x_4 \vee x_5) && \wedge && (x_4 \vee \neg x_5) && \wedge \\
&(\neg x_0 \vee x_2) && \wedge && (x_0 \vee \neg x_4) && \wedge && (\neg x_0 \vee x_7) && \wedge && (\neg x_4 \vee \neg x_7) && \wedge && (x_7 \vee x_8)
\end{align*}

\subsection{20 Clauses \label{ssec:20_clause}}
\begin{align*}
    f(x) =
    & (\neg x_6 \vee \neg x_9) && \wedge && (\neg x_1 \vee \neg x_4) && \wedge && (x_0 \vee x_6) && \wedge && (\neg x_6 \vee \neg x_7) && \wedge && (x_2 \vee \neg x_6) && \wedge\\
    & (x_3 \vee x_8) && \wedge && (\neg x_1 \vee x_4) && \wedge && (\neg x_0 \vee x_6) && \wedge && (x_5 \vee \neg x_9) && \wedge && (x_0 \vee \neg x_8) && \wedge\\
    & (x_1 \vee \neg x_8) && \wedge && (x_1 \vee x_8) && \wedge && (x_5 \vee \neg x_7) && \wedge && (x_2 \vee \neg x_7) && \wedge && (\neg x_0 \vee \neg x_5) && \wedge\\
    & (x_6 \vee \neg x_9) && \wedge && (\neg x_0 \vee \neg x_7) && \wedge && (x_0 \vee x_3) && \wedge && (\neg x_1 \vee x_6) && \wedge && (\neg x_0 \vee x_3)
\end{align*}

\subsection{30 Clauses \label{ssec:30_clause}}
\begin{align*}
    f(x) =& (\neg x_3 \vee x_7) && \wedge && (\neg x_4 \vee x_8) && \wedge && (\neg x_3 \vee \neg x_9) && \wedge && (\neg x_6 \vee \neg x_7) && \wedge && (x_1 \vee x_5) && \wedge\\
    & (\neg x_0 \vee x_6) && \wedge && (\neg x_3 \vee \neg x_4) && \wedge && (x_0 \vee x_8) && \wedge && (x_5 \vee x_7) && \wedge && (x_3 \vee \neg x_8) && \wedge\\
    & (x_1 \vee x_8) && \wedge && (\neg x_0 \vee \neg x_6) && \wedge && (x_1 \vee x_2) && \wedge && (x_0 \vee \neg x_1) && \wedge && (\neg x_5 \vee x_9) && \wedge\\
    & (x_4 \vee \neg x_6) && \wedge && (\neg x_2 \vee \neg x_8) && \wedge && (x_8 \vee \neg x_9) && \wedge && (x_7 \vee \neg x_9) && \wedge && (x_1 \vee \neg x_4) && \wedge\\
    & (x_6 \vee \neg x_9) && \wedge && (x_3 \vee x_4) && \wedge && (\neg x_5 \vee x_6) && \wedge && (x_1 \vee \neg x_9) && \wedge && (x_1 \vee \neg x_3) && \wedge\\
    & (x_2 \vee \neg x_5) && \wedge && (\neg x_0 \vee x_7) && \wedge && (x_0 \vee x_2) && \wedge && (\neg x_0 \vee \neg x_1) && \wedge && (\neg x_7 \vee x_9)
\end{align*}

\twocolumngrid

\bibliography{references}

\end{document}